# TumorHoPe2: An updated database for Tumor Homing Peptides

Diksha Kashyap[1], Devanshi Gupta[1], Naman Kumar Mehta[1], and Gajendra P. S. Raghava[1, *]

1. Department of Computational Biology, Indraprastha Institute of Information Technology, Okhla Phase 3, New Delhi-110020, India.

**Mailing Address of Authors**

Diksha Kashyap (DK): diksha23228@iiitd.ac.in     ORCID ID: https://orcid.org/0009-0009-4211-9636

Devanshi Gupta (DG):  devanshi23227@iiitd.ac.in     ORCID ID: https://orcid.org/0000-0001-6116-6424

Naman Kumar Mehta (NKM):   namanm@iiitd.ac.in     ORCID ID: https://orcid.org/0009-0009-0244-2826

Gajendra P. S. Raghava (GPSR): raghava@iiitd.ac.in     ORCID ID: https://orcid.org/0000-0002-8902-2876

**\*Corresponding Author**

Prof. Gajendra P. S. Raghava

Head and Professor

Department of Computational Biology

Indraprastha Institute of Information Technology, Delhi

Okhla Industrial Estate, Phase III, (Near Govind Puri Metro Station)

New Delhi, India – 110020

Office: A-302 (R&D Block)

Phone: 011-26907444

Email: raghava@iiitd.ac.in

Website: http://webs.iiitd.edu.in/raghava/

# Abstract

Addressing the growing need for organized data on tumor homing peptides (THPs), we present TumorHoPe2, a manually curated database offering extensive details on experimentally validated THPs. This represents a significant update to TumorHoPe, originally developed by our group in 2012. TumorHoPe2 now contains 1847 entries, representing 1297 unique tumor homing peptides, a substantial expansion from the 744 entries in its predecessor. For each peptide, the database provides critical information, including sequence, terminal or chemical modifications, corresponding cancer cell lines, and specific tumor types targeted. The database compiles data from two primary sources: phage display libraries, which are commonly used to identify peptide ligands targeting tumor-specific markers, and synthetic peptides, which are chemically modified to enhance properties such as stability, binding affinity, and specificity. Our dataset includes 594 chemically modified peptides, with 255 having N-terminal and 195 C-terminal modifications. These THPs have been validated against 172 cancer cell lines and demonstrate specificity for 37 distinct tumor types. To maximize utility for the research community, TumorHoPe2 is equipped with intuitive tools for data searching, filtering, and analysis, alongside a RESTful API for efficient programmatic access and integration into bioinformatics pipelines. It is freely available at https://webs.iiitd.edu.in/raghava/tumorhope2/

# Author's Biograph

1. Diksha Kashyap is currently pursuing a Master's degree in Computational Biology at the Department of Computational Biology, Indraprastha Institute of Information Technology, New Delhi, India.
2. Devanshi Gupta is currently pursuing a Master's degree in Computational Biology at the Department of Computational Biology, Indraprastha Institute of Information Technology, New Delhi, India.
3. Naman Kumar Mehta is currently working as Ph.D. in Computational biology from Department of Computational Biology, Indraprastha Institute of Information Technology, New Delhi, India
4. Gajendra P. S. Raghava is currently working as Professor and Head of Department of Computational Biology, Indraprastha Institute of Information Technology, New Delhi, India.

# Introduction

Cancer is one of the serious global health challenges of the 21st century, affecting millions of lives each year. According to recent estimates, one in five people worldwide will develop cancer during their lifespan [1]. While advances in early detection and prevention have helped reduce cancer mortality, developing effective treatments remains a significant challenge. Traditional chemotherapy and radiation therapies often lack selectivity, damaging both cancerous and healthy cells, causing severe side effects and drug resistance [2]. Additionally, the emergence of drug resistance in cancer cells further complicates treatment efforts, highlighting the urgent need for more selective and effective therapeutic strategies. In recent years, targeted drug delivery systems have emerged as a promising approach to overcoming these challenges [3]. Among these, tumor-homing peptides (THPs) have gained significant attention due to their ability to selectively bind to tumor cells and tumor vasculature [4]. These short peptide sequences can recognize cancer-associated markers that are either uniquely expressed or overexpressed on tumor cells, allowing for precise targeting. THPs have been widely utilized for drug delivery, tumor imaging, and nanomedicine applications, enhancing the therapeutic efficacy of anti-cancer agents while reducing systemic toxicity [5].

Tumor-homing peptides (THPs) represent a promising class of molecules for targeted cancer therapy and diagnostics. One of the most powerful techniques for their discovery is phage display technology, which enables high-throughput screening of peptides that bind to specific tumor markers [6]. Over the years, numerous THPs have been identified and extensively tested in preclinical and clinical settings. Many of these peptides have demonstrated remarkable tumor-targeting capabilities, significantly enhancing drug accumulation at the tumor site when conjugated to therapeutic agents. Beyond drug delivery, THPs have also proven valuable in transporting imaging agents, thereby improving early tumor detection and disease monitoring [7]. However, despite their considerable potential, information on THPs remains fragmented across various scientific studies, lacking a centralized resource for systematic compilation and organization. To address this critical challenge, our group developed a pioneering database, TumorHoPe, in 2012, which maintains comprehensive information on experimentally validated tumor-homing peptides [5]. This repository offers detailed information on peptide sequences, target receptors, tumor specificity, experimental validation methods, and potential therapeutic applications. By integrating curated data, structural insights, and functional annotations, TumorHoPe has served as a valuable resource for researchers engaged in targeted cancer therapies and precision medicine [6–8]. The utility of this database is further evidenced by its widespread use and numerous citations within the scientific community over the years. Furthermore, datasets derived from TumorHoPe have facilitated the development of a diverse range of computational methods for predicting tumor-homing peptides, including TumorHPD, THPep, NEPTUNE, StackTHPred, LLM4THP, and SCMTHP [4,9–16]. In essence, TumorHoPe has been a cornerstone resource for researchers in the field of targeted drug discovery against cancer.

To the best of our knowledge, TumorHoPe remains the sole comprehensive resource cataloging detailed information on tumor-homing peptides. Developed in 2012, this database has not undergone any updates for the past twelve years, and no alternative or comparable database has emerged to fill this void. Despite the recognized importance and exponential growth in the field of THPs, there is a critical absence of a centralized repository that systematically maintains information on newly discovered THPs. This significant gap underscores the urgent need for an updated and expanded resource to support ongoing research and development in targeted cancer therapies. To address this critical challenge, we have updated our original repository, TumorHoPe, to create TumorHoPe2, which now maintains comprehensive and updated information up to the year 2024. Our revised repository, TumorHoPe2, contains more than double the entries found in the original version of TumorHoPe. Furthermore, a suite of enhanced functionalities has been integrated into our web interface to facilitate user interaction and data exploration. We anticipate that this updated database will prove to be a significant step forward in leveraging bioinformatics to accelerate the discovery and development of next-generation cancer therapeutics.

## Material and Methodology

### Data collection

The updated TumorHoPe2 database comprises 1,847 entries, including 487 from PubMed, 652 from Patent Lens, and 744 from the previous TumorHoPe version; among these, 6 sequences were overlapping. Data were systematically curated from reliable sources covering studies published between 2012 and 2024. Targeted keywords like 'tumor-homing peptides' and 'tumor-targeting peptides' were used to identify relevant research. All entries consist of manually curated, experimentally validated peptide sequences extracted from peer-reviewed articles and patents.

### Database Architecture

TumorHoPe2 is built on an updated technology stack to ensure better performance, scalability, and user experience. The database is powered by Apache HTTP Server, with MySQL as the backend for efficient data storage and retrieval. The front end is developed using PHP, HTML, and JavaScript, providing an interactive and user-friendly interface. The choice of open-source and platform-independent technologies ensures accessibility and flexibility for researchers worldwide. The updated architecture of TumorHoPe2 is illustrated in (Figure 1).

### Organization of data

TumorHoPe2 is a manually curated updated database that systematically compiles information on peptides capable of targeting, binding to, or homing to tumors. The data for each peptide is categorized into primary and secondary information. Primary data includes essential details such as the peptide sequence, PubMed ID (PMID), and experimental validation. Secondary data encompasses additional properties, including secondary and tertiary structures, amino acid composition, frequency, and physicochemical characteristics. Each peptide is assigned a unique entry number, and the information is structured across multiple tables, with each table dedicated to specific aspects of the peptide. This structured organization ensures that researchers can efficiently retrieve and analyze relevant data on tumor-homing peptides.

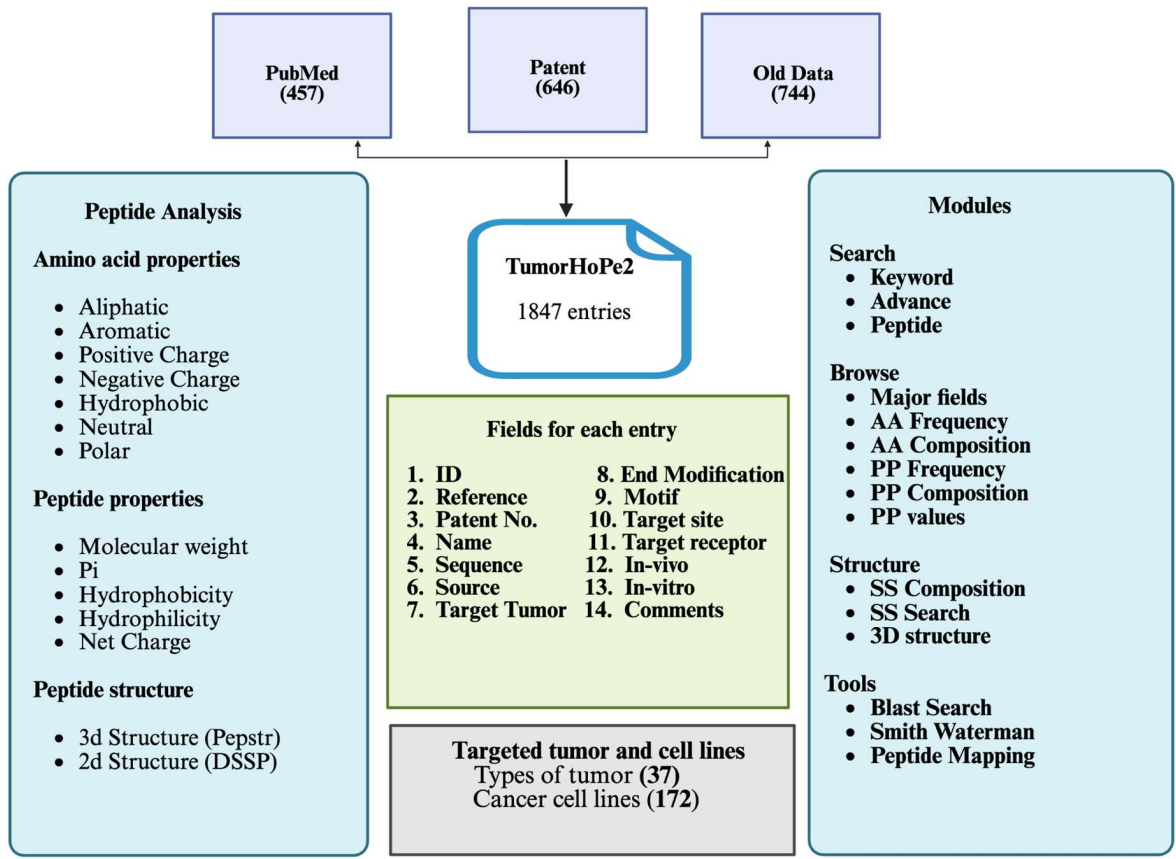

**Figure 1:** Architecture of TumorHoPe2 database.

## Web interface and tools implementation

### Search Tools

TumorHoPe2 integrates a suite of enhanced search tools designed to enable efficient and flexible retrieval of tumor-homing peptide data. Users can explore the database through keyword, advanced, and peptide-based search options. For general exploration, the keyword search allows simple input of any relevant term or phrase to scan across all database fields. The advanced search interface enables users to build complex queries by selecting specific fields such as peptide sequence, motif, or target tumor, defining search terms, and applying logical operators like AND, OR, and NOT. This interface supports stepwise query construction and allows real-time review and editing of the full query. Sequence-based retrieval in TumorHoPe2 is facilitated through the peptide search module, which offers two search modes: a subsequence search that identifies peptides containing the input fragment, and an exact match search that returns peptides identical to the query sequence. These updated search functionalities make TumorHoPe2 more intuitive and versatile while preserving the core structure and intent of the earlier version.

## Browse Tools

A versatile browsing interface facilitates comprehensive exploration of tumor-homing peptides based on various biological and chemical properties. Users can filter peptides by key parameters such as target tumor type, cancer cell lines, year of publication, or target site, enabling detailed analysis within specific experimental and clinical contexts. The platform supports searches based on amino acid frequency and overall composition, allowing precise retrieval of peptides with defined residue counts or characteristic amino acid distributions. In addition, filtering by physicochemical properties—including charge, polarity, hydrophobicity, net charge, molecular weight, and isoelectric point—provides valuable means to identify peptides with features critical for tumor targeting and therapeutic applications.

## Structure-Based Retrieval

We incorporated predicted secondary and tertiary structures to enable structural analysis of tumor-homing peptides. Using the Secondary Structure Composition tool, users can browse peptides based on the proportion of structural elements defined by DSSP—helix, strand, turn, and coil. The Secondary Structure Search interface allows retrieval of peptides with specific structural patterns. An interactive 3D viewer built with NGL is used to display peptide tertiary structures, offering options to adjust visualization styles and orientations [17].

## Analysis tools

We implemented sequence similarity searches using BLAST and Smith–Waterman algorithms to enable precise alignment of user-submitted peptides against the database [18,19]. Local similarity is evaluated through sliding window alignment to identify regions with the highest number of identical residues. The peptide mapping feature allows users to search for peptides within larger protein sequences or find database peptides related to query proteins. Additionally, the platform calculates essential physicochemical properties such as hydrophobicity, net charge, isoelectric point, and molecular weight for peptides submitted in FASTA format.

## REST API Support

The updated TumorHoPe2 supports programmatic data retrieval, reducing the reliance on manual downloads. Users can create query links in URL, CURL, or Wget formats, with results returned in JSON. The REST API supports retrieval of tumor-homing peptide data based on fields like Target Cell or Source, enabling smooth integration into research workflows and applications.

## Results and Discussion

The TumorHoPe2 database has been systematically updated to incorporate an additional 1,103 tumor-homing peptides (THPs) curated from two primary sources: PubMed and Patent Lens. Specifically, 652 peptides were obtained from Patent Lens and 457 from PubMed, with 6 peptides overlapping between the two sources. These newly curated entries were integrated with 744 peptides from the previous version of TumorHoPe, bringing the total number of unique peptides in the updated database to 1,847 The detailed length distribution of the number of unique tumor-homing peptides is provided in (**Figure 2**).

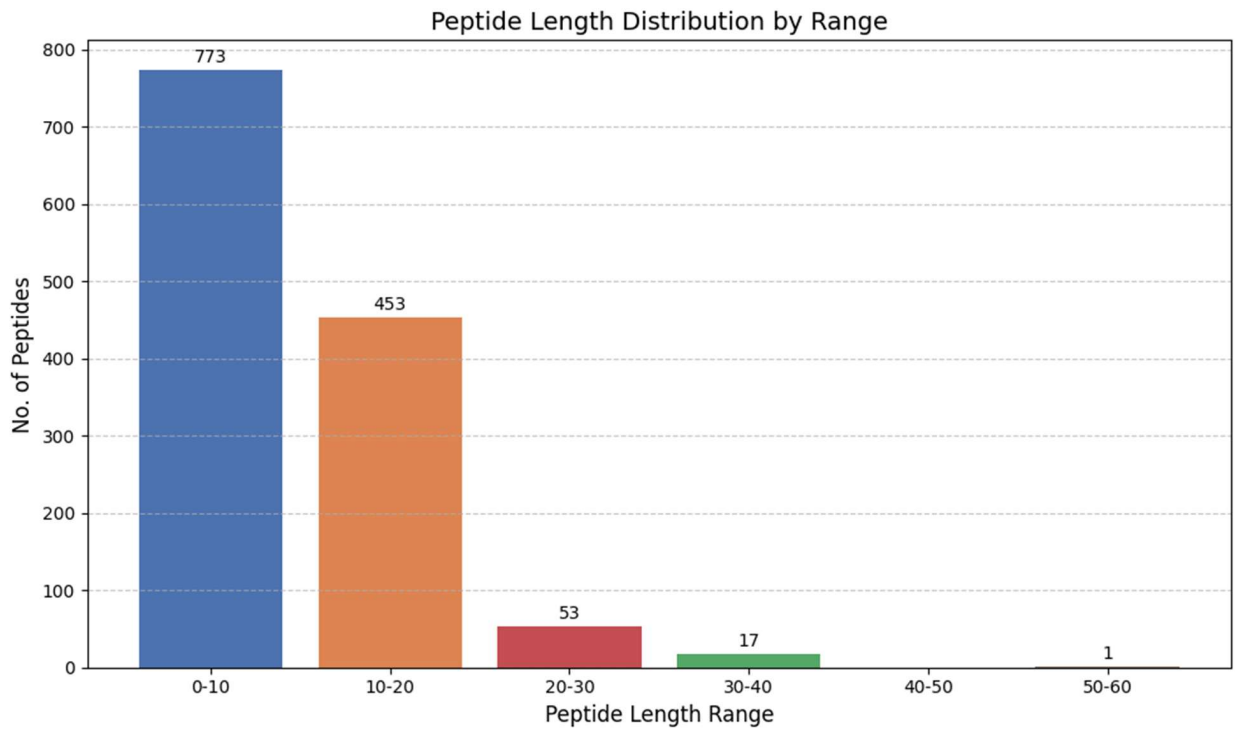

**Figure 2:** Length wise distribution of the number of unique tumor-homing peptides.

The distribution shows that peptides of lengths 7, 9, 10, and 12 amino acids were the most frequent, with 9-residue peptides being the most abundant. Peptides with lengths greater than **20** amino acids were relatively rare.

The newly curated dataset spans a diverse range of cancer types, with a significant proportion of peptides associated with breast cancer (511), followed by lung cancer (151), prostate cancer (146), melanoma (135), and brain cancer (134). Other notable cancer types include pancreatic cancer (94), colon cancer (58), gastric cancer (54), and medullary thyroid (33). The database also includes peptides targeting less frequently studied cancers such as osteosarcoma (7), human neuroblastoma (7), and esophagus and gastroesophageal cancer (8). The distribution of the number of entries of peptide across different cancer types is illustrated in (Figure 3).

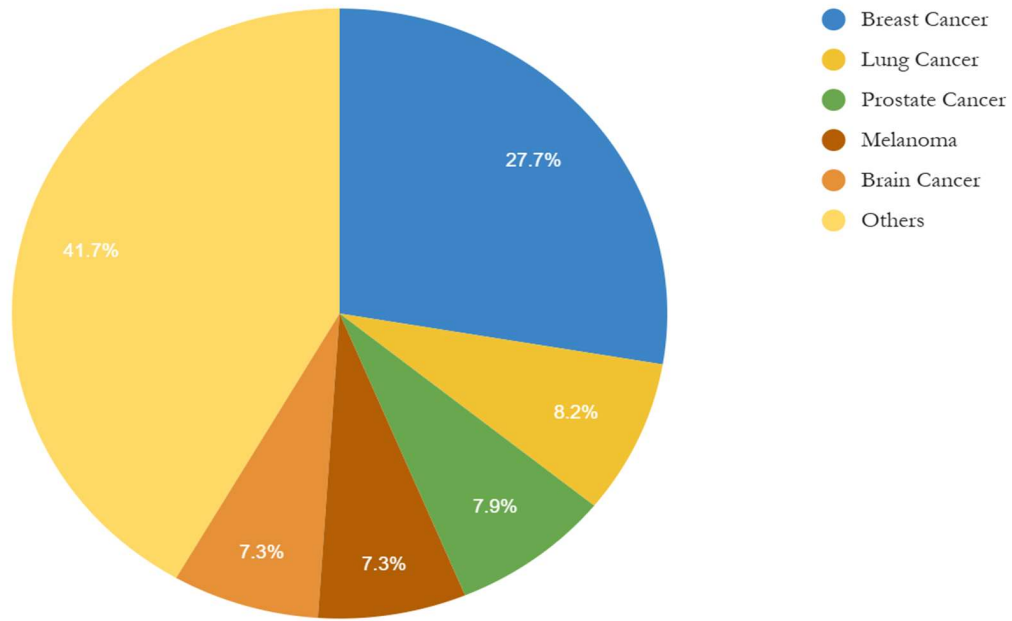

**Figure 3:** Distribution of number of entries of peptide across different cancer types.

The updated dataset includes peptides validated in various cancer cell lines, with the highest number of peptides associated with MDA-MB-435 (292 peptides), 4T1 (233), B16F1 (214), PPC1 (206), and PC-3 (209). Additionally, a considerable number of peptides were tested on MCF-7 (146), HeLa (71) and A549 (55) cell lines.

The distribution of cancer cell lines within the database's entries, across the frequency of their occurrence across all data points, is illustrated in (Figure 4).

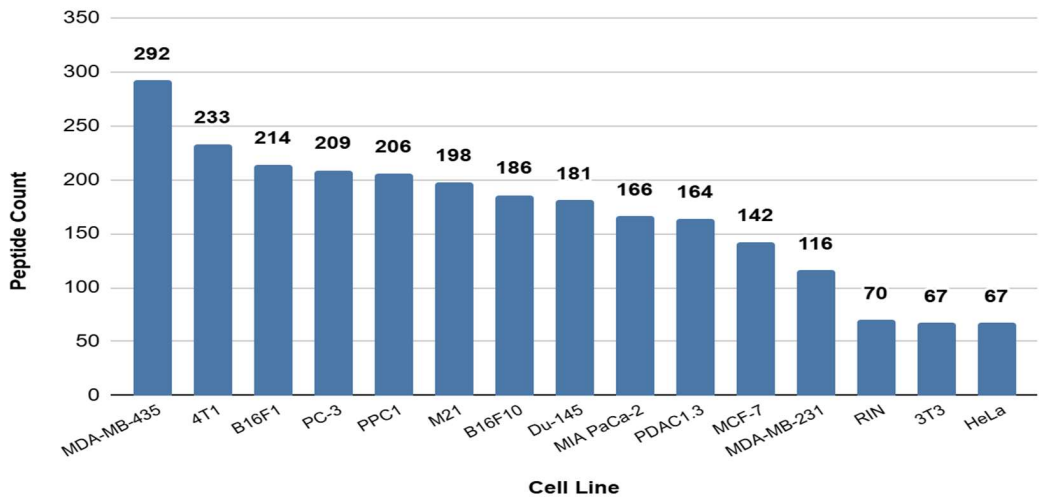

**Figure 4:** Distribution of peptides validated in various cancer cell line

One of the significant improvements in TumorHoPe2 is the enhanced annotation of peptides, including detailed structural and physicochemical properties. The addition of secondary and tertiary structure predictions allows researchers to understand better how these peptides interact with tumor cells, paving the way for more precise drug development. The dataset contains 1,789 non-cyclic peptides, 58 cyclic peptides, and one peptide with both linear and cyclic forms. This classification provides insight into the structural preferences of tumor-homing peptides. We analyzed the secondary structure distribution of tumor-homing peptides to provide insights into their structural properties. The distribution of the number of cyclic and non-cyclic peptide is reported in (Figure 5).

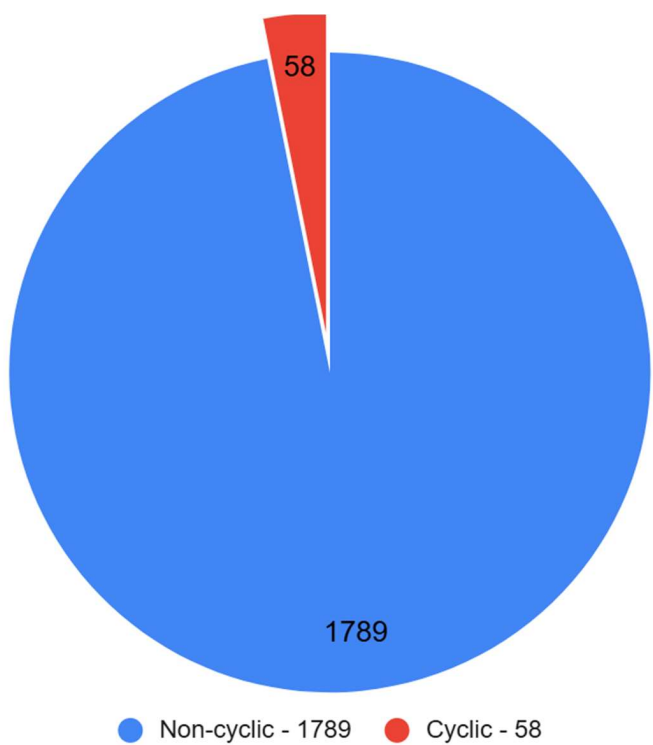

**Figure 5:** Distribution of the number of entries of cyclic and non-cyclic peptides

Searching for relevant peptides has also become easier, with an improved interface that supports keyword, peptide, and advanced search options, making data retrieval more intuitive and efficient.

Among the curated peptides, we identified several well-established tumor-homing motifs. The most prevalent motifs include RGD (89 peptides), NGR (54 peptides), and CendR (51 peptides). Other motifs, such as RXD/QXR (8 peptides), LTVXPW (3 peptides), and bombesin (2 peptides), were also documented.

Beyond just compiling data, TumorHoPe2 serves as a bridge between research and application. Tumor-homing peptides have shown great promise in drug delivery, cancer imaging, and targeted therapy. But finding detailed information about these peptides hasn't been easy—it's usually spread out across different sources, making it hard for researchers to get everything they need in one place. TumorHoPe2 addresses this challenge by providing a well-organized, easily accessible interface that is tailored to the requirements of the scientific community.

A majority (1,608 peptides) have low helical content, while 1,790 lack significant beta strands, and 709 show a low frequency of turns, as shown in Table 1. Peptides with moderate structural content are less common, with 136 containing helices, 46 exhibiting beta strands, and 630 showing a balanced turn composition. A small fraction (85 peptides) displays increased helicity, while only 7 have notable beta strands, and 470 contain moderate turns. Peptides with high structural content are rare, with just 14 having prominent helices, no peptides exhibiting beta strands, and 34 showing significant turns. Notably, 495 peptides are predominantly composed of coil structures, reinforcing the inherent structural flexibility of short peptides.

**Table 1:** Detailed distribution of different types of secondary structures

| Structural Content (%) | Helical Peptides | Beta Strand Peptides | Turn Peptides |
|---|---|---|---|
| 0-20% | 1608 | 1790 | 709 |
| 20-40% | 136 | 46 | 630 |
| 40-60% | 85 | 7 | 470 |
| 60-80% | 14 | 0 | 34 |
| 80-100% | 0 | 0 | 495 |

While this update marks a significant step forward, there is still room for growth. Experimental validation methods for tumor-homing peptides vary, and there is a need for more standardised datasets to further improve the reliability of predictions. Future updates will focus on integrating predictive modeling tools, expanding the dataset with newly published peptides, and enhancing the platform's capabilities for screening and designing novel peptides. By continuously evolving, TumorHoPe2 aims to support researchers in the ongoing fight against cancer, helping to accelerate discoveries that bring targeted therapies closer to clinical application.

## Comparison with previous version

TumorHoPe2 represents a significant advancement over the original version, with notable increases in both the quantity and diversity of peptide data. The total number of entries has grown from 744 to 1,847, and the number of unique peptides has expanded from 707 to 1,297, reflecting the integration of newly identified sequences along with refined annotations. The updated version includes data associated with over 172 cell lines, compared to just over 80 previously, offering a broader experimental framework. Likewise, the number of target tumor types has been extended to cover more than 30 distinct tumors, enhancing the applicability of the database across various cancer contexts. In terms of peptide composition, the number of natural sequences has more than doubled, increasing from 571 to 1,253, while chemically modified peptides have risen from 173 to 594, underscoring a growing emphasis on engineered peptides with therapeutic and diagnostic relevance. Additionally, TumorHoPe2 incorporates the MAP (Modification and Annotation in Proteins) format, which enables the embedding of tags directly within peptide sequences to denote residue-level modifications. These annotations capture details such as chemical modifications, incorporation of non-standard amino acids, binding sites, and sequence mutations, facilitating more precise structural and functional analyses [20]. (Figure 6) provides a visual representation, highlighting the advancements in TumorHoPe2.0 compared to its previous version.

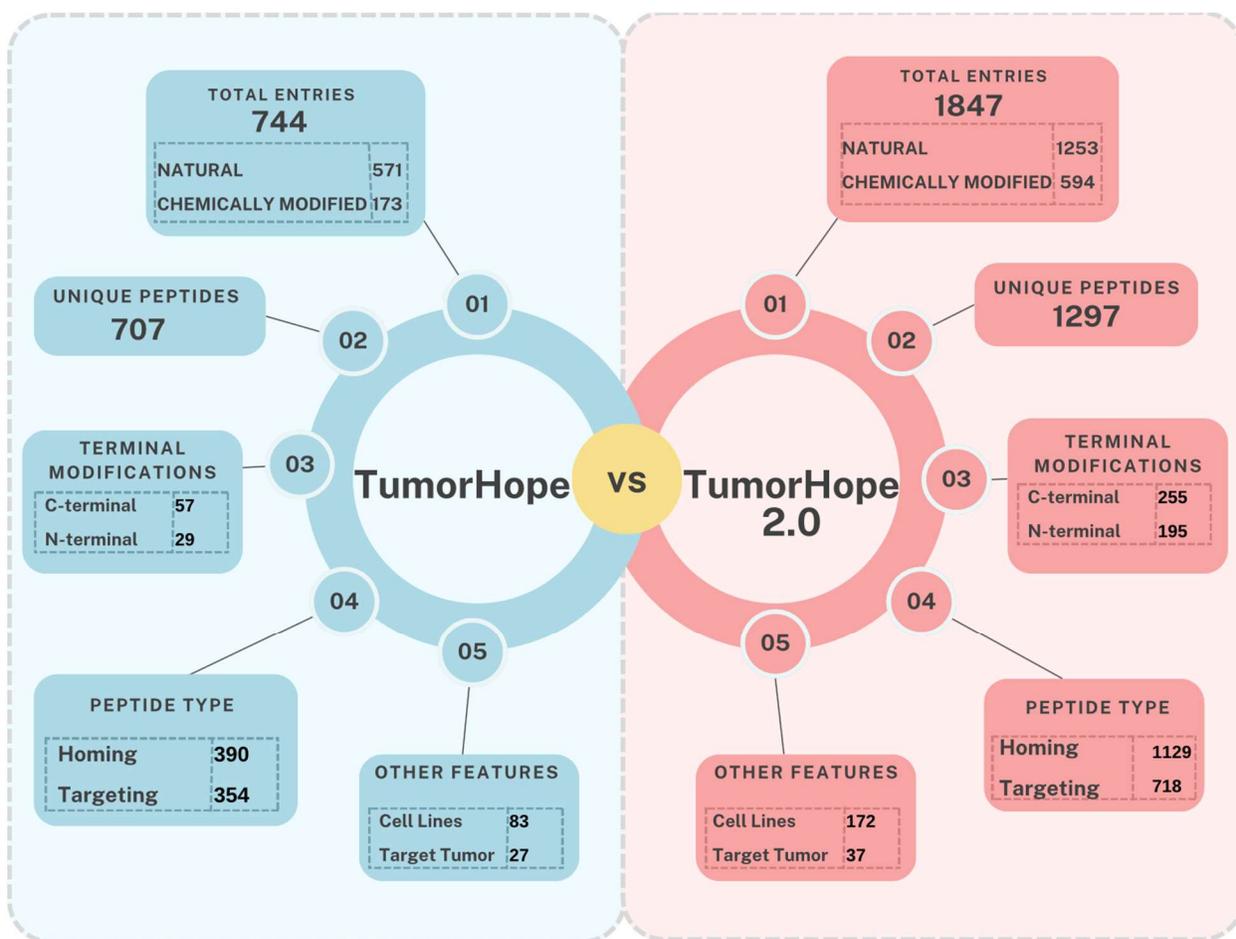

**Figure 6:** Visual representation highlighting the advancements in TumorHoPe2.0

The complete comparison entries between TumorHoPe and TumorHoPe2 are provided in Table 2.

**Table 2:** Comparison of entries between TumorHoPe and TumorHoPe2

| Keyword | TumorHope | TumorHope2 |
|---|---:|---:|
| Total entries | 744 | 1847 |
| Unique peptides | 707 | 1297 |
| Homing Peptides | 390 | 1129 |
| Targeting Peptides | 354 | 718 |
| Cell lines | 83 | 172 |
| Target Tumor | 27 | 37 |
| N-terminal modifications | 57 | 255 |
| C-terminal modifications | 29 | 195 |
| Natural sequences | 571 | 1253 |
| Chemically modified sequences | 173 | 594 |

## Limitation and update of TumorHoPe2

The TumorHoPe2 database is regularly updated to incorporate newly reported tumor-homing peptides from the scientific literature and verified sources. These updates enhance the accuracy, consistency, and reliability of the dataset by refining peptide annotations, updating experimental details, and integrating improved search and browsing functionalities. However, our current capabilities do not allow for precise prediction of the tertiary structures of highly complex modified peptides, such as those containing tert-butyloxycarbonyl (Boc), hydroxyproline, norleucine, nitrile groups, or carbodiimide-mediated amines, due to the unavailability of a force field library.

## Availability

The TumorHoPe2 database, a comprehensive resource for tumor-homing peptides and related information, is freely accessible to the public at https://webs.iiitd.edu.in/raghava/tumorhope2/. Users can access and download the complete dataset through the Downloads section of the website, which offers files in multiple formats for ease of use in computational and experimental research.

The API can be accessed and configured through the interface available at https://webs.iiitd.edu.in/raghava/tumorhope2/restapi/Tumorhope_api.php, enabling seamless integration with bioinformatics pipelines and automated workflows.

## Funding Source

The current work has been supported by the Department of Biotechnology (DBT) grant BT/PR40158/BTIS/137/24/2021.

## Conflict of interest

The authors declare no competing financial and non-financial interests.

## Authors' contributions

DK, DG, and NKM manually collected the data. DK, DG, and NKM manually curated and analysed the data. DK, DG, developed the backend and frontend of the webserver. DK, DG, NKM and GPSR prepared the manuscript. DK, DG, NKM, and GPSR. reviewed the manuscript. GPSR conceived and coordinated the project. All authors read and approved the final manuscript.

## Acknowledgments

Authors are thankful to the University Grants Commission (UGC), Department of Science and Technology (DST-INSPIRE), for fellowships and financial support, Indraprastha Institute of Information Technology (IIITD), for fellowships and financial support, and the Department of Computational Biology, IIITD, New Delhi, for infrastructure and facilities. We would like to acknowledge that Figures were created using BioRender.com.

for representing chemical modifications, annotations, and mutations in protein sequences: An extension of the FASTA format, arXiv [q-bio.BM] (2025). https://doi.org/10.48550/ARXIV.2505.03403.